\documentclass{article}
\usepackage{spconf,amsmath,graphicx}
\usepackage{bbm}
\usepackage{multirow}
\usepackage{hyperref}

\def\x{{\mathbf x}}
\def\y{{\mathbf y}}

\def\causale{{\mathbf e}^{s}}
\def\noncausale{{\mathbf e}^{a}}

\usepackage{xcolor}
\usepackage{tipa}
\definecolor{darkspringgreen}{rgb}{0.09, 0.45, 0.27}

\title{Cascaded encoders for unifying streaming and non-streaming ASR}
\name{
\begin{tabular}{c}
Arun Narayanan, Tara N. Sainath, Ruoming Pang, Jiahui Yu, Chung-Cheng Chiu,\\
Rohit Prabhavalkar, Ehsan Variani, Trevor Strohman
\end{tabular}
}
\address{Google LLC, USA \\
\fontsize{9}{9}\selectfont\ttfamily\upshape
\{arunnt\}@google.com}
\begin{document}
\ninept
\maketitle
\begin{abstract}
End-to-end (E2E) automatic speech recognition (ASR) models, by now, have shown competitive performance on several benchmarks. These models are structured to either operate in streaming or non-streaming mode. This work presents \emph{cascaded encoders} for building a single E2E ASR model that can operate in both these modes simultaneously. The proposed model consists of streaming and non-streaming encoders. Input features are first processed by the streaming encoder; the non-streaming encoder operates exclusively on the output of the streaming encoder. A single decoder then learns to decode either using the output of the streaming or the non-streaming encoder. Results show that this model achieves similar word error rates (WER) as a standalone streaming model when operating in streaming mode, and obtains 10\% -- 27\% relative improvement when operating in non-streaming mode. Our results also show that the proposed approach outperforms existing E2E two-pass models, especially on long-form speech. 
\end{abstract}
\section{Introduction} \label{sec:intro}
\vspace{-0.05in}
End-to-end (E2E) automatic speech recognition systems (ASR) have made tremendous progress over the last few years, achieving word error rates (WER) that match or surpass conventional ASR models in several common benchmarks \cite{park2019specaugment, gulati2020conformer, sainath2020streaming, li2020comparison}. Typical E2E systems consist of a single neural network that transforms input audio to sequences of output tokens, like characters or word-pieces, that can be readily transformed to the final sequence of words. Examples of such models include connectionist temporal classification (CTC) \cite{graves2006connectionist}, attention-based encoder-decoder models \cite{ChorowskiBahdanauSerdyukChoEtAl15} like listen-attend-spell (LAS) \cite{chan2016listen}, recurrent neural net transducer (RNN-T) \cite{Graves12}, and other interesting variations \cite{kim2017joint, zhang2020transformer, yeh2019transformer}.

The architecture of E2E model, and importantly that of the encoder, partly depends on the target application that the model will be used for. For instance, a number of applications that involve end-user interaction, like voice-search or on-device dictation, require the model to perform recognition in a streaming fashion\footnote{In \emph{streaming} ASR, the words are expected to be output as they are spoken, with as little latency as possible.}. As a result, models that use future context to improve performance, like the ones that use bi-directional LSTMs, are not suitable. On the other hand, applications like offline video captioning do not require the model to be streaming and can make use of future context to improve performance. Since these requirements are task specific, separate models are usually trained for streaming and non-streaming applications.

For streaming applications, two-pass models have been proposed to make-up for the performance loss due to lack of future context \cite{sainath2019two,hu2020deliberation}. Similar to a conventional second pass rescoring model \cite{hori2007efficient}, these models are used to rescore hypotheses generated by the streaming first pass model. For example, Sainath {\it et al.} \cite{sainath2019two} use an LAS decoder that attends on the output of the encoder of a streaming RNN-T model. Hu {\it et al.} extend this work by using a deliberation model \cite{xia2017deliberation} that simultaneously attends on the output of the encoder and the first-pass hypotheses. Apart from rescoring the first pass hypotheses, the second pass models can also be used to re-decode the input at the expense of additional computation cost.

One of the shortcomings of LAS decoder based two-pass models is that they suffer from long-form issues -- the word deletion rate of an LAS decoder is higher compared to an RNN-T decoder on long-form speech \cite{chiu2019comparison}. Therefore, LAS-based two-pass models are not applicable in long-form conditions, as we show in our experiments. The model that we propose, called cascaded encoders, alleviates this problem, and enables the same model to be used for streaming and non-streaming applications with improved performance in all conditions. It does that by replacing the second pass LAS decoder with an RNN-T decoder that is shared with the first pass model.

Cascaded encoders is motivated by the observation that a single model can behave differently based on the input it receives \cite{dabre2019recurrent,shi2020sequence}. Shi {\it et al.} \cite{shi2020sequence} propose a multi-talker ASR model that predicts targets in an iterative fashion by changing the features that are input to a shared model. Dabre and Fujita \cite{dabre2019recurrent} use the same parameters for all layers of an encoder, and show that the recurrence depth can be dynamically varied during inference. The main observation is that a single model can learn to map different input distributions to the same or different targets, if they have been exposed to these mappings during training. The work on multi-domain models for ASR \cite{NarayananMisraSimPundakEtAl18} and domain (or dialect) identifier-based multidomain \cite{sainath2020streaming} and multilingual \cite{li2018multi} models also exploit this behaviour. 

With cascaded encoders, we show that a single RNN-T decoder can learn the distribution of higher-level acoustic features generated by both streaming and non-streaming encoders of an E2E model. Furthermore, we show that the non-streaming encoder can be trained directly on the output of the streaming encoder instead of input acoustic features; this allows the non-streaming decoder to use fewer layers as opposed to a fully non-streaming model, with a very limited loss in performance. The idea of cascaded encoders is most similar to the recent works on universal ASR \cite{anonymous2021universal} and transformer-transducer \cite{tripathi2020ttone} for unifying streaming and non-streaming models. While these works focus on specific encoder architectures (conformer \cite{gulati2020conformer} and context-net \cite{han2020contextnet} in \cite{anonymous2021universal}, and tranformers in \cite{tripathi2020ttone}), the current work shows that any pairing of encoders can be used to unify streaming and non-streaming use-cases.

The rest of the paper is organized as follows. In Sec.~\ref{sec:model} we discuss the model architecture and the methodology for training cascaded encoders. We present our experimental setup in Sec.~\ref{sec:experiments}. Sec.~\ref{sec:results} presents results and comparisons that show performance using various combinations of architectures, and in a latency constrained setting. Conclusions are presented in Sec.~\ref{sec:conclusion}. 
\section{Model description} \label{sec:model}
\begin{figure}[h!]
  \centering
  \includegraphics[scale=0.5]{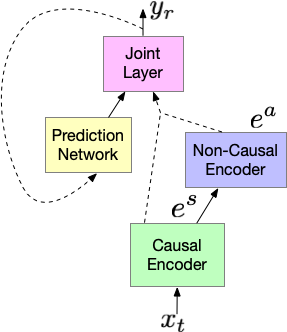}
  \caption{{A block diagram of the cascaded encoders model.}}
   \label{fig:cascade}
   \vspace{-0.1in}
\end{figure}

\vspace{-0.05in}
The block diagram of the proposed model is shown in Fig.~\ref{fig:cascade}. The model is based on the recurrent neural net transducer (RNN-T), which consists of an encoder and a decoder. As input, the encoder of an RNN-T receives acoustic features,  $\x = [\x_1, \cdots, \x_{T_x}]$, where $\x_t \in \mathbf{R}^d$ corresponds to the input at time frame, $t$. Unlike attention based E2E models, like LAS \cite{chan2016listen}, the RNN-T decoder has a prediction network, typically an LSTM, that directly models the label sequence $\y = [\y_1, \cdots, \y_R]$, where $\y_r \in \mathcal{Y}$, the output token vocabulary. The prediction network processes labels independent of the acoustics; the RNN-T decoder uses a joint network to combine acoustic encoder's output and the prediction network's output to make the final prediction. 

\subsection{Cascaded Encoders}
The encoder layers of an RNN-T, typically, is either causal, like a uni-directional LSTM, or non-causal, like a bidirectional LSTM\footnote{We use causal / streaming and non-causal / non-streaming interchangeably in this work.}. But the proposed cascaded encoders model consists of both causal and non-causal layers. The input features, $\x$, are first passed to a causal encoder, which transforms the features to a higher-level representation, $\causale = [\causale_1, \cdots, \causale_{T_e}]$. The non-causal encoder, which is connected in \emph{cascade} to the causal encoder, receives $\causale$ as input, and outputs $\noncausale = [\noncausale_1, \cdots, \noncausale_{T_e}]$. Both the causal and the non-causal encoders are directly connected to a shared RNN-T decoder.

Since there are 2 input processing paths -- [casual-encoder $\rightarrow$ RNN-T decoder] and [causal-encoder $\rightarrow$ non-causal-encoder $\rightarrow$ RNN-T decoder] -- the model's loss consist of 2 components:
\begin{align}
\mathcal{L}_s
              &= -\sum_{\{(\x \rightarrow \causale, \y)\}} \log P(\y|\causale). \\
\mathcal{L}_a &= -\sum_{\{(\x \rightarrow \noncausale, \y)\}} \log P(\y|\noncausale),
\end{align}
where $\mathcal{L}_s$ and $\mathcal{L}_a$ are the losses from the causal and the non-causal connections to the decoder, respectively. The total loss can be computed as the weighted sum of these components:
\begin{equation}
    \mathcal{L} = \lambda \mathcal{L}_s + (1 - \lambda) \mathcal{L}_a,
\end{equation}
where $\lambda$ is a weighting term. In practice, we found that we can decrease the step-time during training by sampling from $\causale$ and $\noncausale$ within a mini-batch using $\lambda$ as the sampling rate. Therefore, for each input utterance, we stochastically choose either the causal or the non-causal processing path at each training step. This alleviates the need to compute RNN-T loss twice for each training example at each training step. With sampling, the model converges roughly after the same number of steps as a standalone streaming model.

During inference, the model can operate either in causal or non-causal mode, depending on the output encoder from which the features are fetched for decoding.

Cascaded encoders offer a few advantages over the LAS-based two-pass models. Training such models, typically, is a two-stage process, since the first and the second pass models are trained independently. But cascaded encoders can be trained in a single stage, simplifying the overall process. The model also use a shared RNN-T decoder, which is better suited to handle long-form speech than LAS \cite{chiu2019comparison}: LAS-based second pass models do not provide improvements over the first pass model in long-form conditions; it can only operate in rescoring mode when used with long-form audio. Cascaded encoders, on the other hand, can provide gains in all conditions, as we show in Sec.~\ref{sec:results}. A potential disadvantage of cascaded encoders is that it is more expensive to use the non-causal layers to do rescoring compared to an LAS-based model, because the RNN-T decoder has to marginalize over all possible alignments to compute the posterior of an output sequence that needs rescoring. 

\subsection{Encoder architectures}
Apart from the widely used LSTM model, we also present results using the conformer architecture \cite{gulati2020conformer} in Sec.~\ref{sec:results}. Conformer combines the best of transformer \cite{vaswani2017attention} and context-net models \cite{han2020contextnet}; it uses both multi-head self attention and convolutional modules within each layer to efficiently combine local and global information. Since conformer is based on transformers, it is also straightforward to limit the amount of right context used in each conformer layer, and hence the entire model, unlike the limited-context LSTM models \cite{zhang2016highway, chen2016training}. This can be done my masking the attention score so that only a window of contiguous frames are used to compute the output, instead of the entire input \cite{zhang2020transformer}. The same strategy can be used to replace LSTMs with a fully causal conformer encoder in an RNN-T model, and referred to as conformer-transducer (C-T) in this work, to differentiate it from RNN-T.

In this work, we examine using conformer layers to implement either the causal encoder or the non-causal encoder or both, and show that encoders can be cascaded irrespective of the architecture adopted to implement the causal and non-causal layers of the model.

\subsection{FastEmit} \label{sec:fastemit}
For streaming applications, it important to make predictions with minimum per-word and endpointer latency. Merely using a causal encoder is not always sufficient to meet these constraints \cite{li2020rnntep}. Additional penalties are typically added to the RNN-T loss to improve overall latency, at the expense of increased WER. In this work, we explore whether the losses in WER as a result of adding latency penalties can be recouped using cascaded encoders.

We use the recently proposed FaseEmit loss to improve overall latency ($\mathcal{L}_{fe}$) \cite{Yu2021fastemit}. With this, the total loss of the model is:
\begin{equation}
    \mathcal{L} = \lambda \mathcal{L}_s + (1 - \lambda) \mathcal{L}_a + \beta \mathcal{L}_{fe} \odot \mathbbm{1}_{causal},
\end{equation}
where, $\beta$ is a tunable hyper-parameter. $\odot$ stands for element-wise multiplication and $\mathbbm{1}_{causal}$ is an indicator function that is $1$ only when the input features to the decoder comes from the causal encoder; $\mathcal{L}_{fe}$ is applied only when the decoder is trained in the causal mode since latency is important mainly for the streaming use-case. When optimizing the model to improve overall latency, we also use minimum WER training \cite{prabhavalkar2018mwer} as it was found to provide better WER-latency trade-offs \cite{li2020rnntep}.

\section{Experimental settings} \label{sec:experiments}
\vspace{-0.05in}

\subsection{Datasets}

Our models are trained on the multidomain training sets described in greater detail in \cite{NarayananMisraSimPundakEtAl18,narayanan2019longform}. The set consists of anonymized and hand-transcribed English utterances that are representative of Google's traffic, including handheld and far-field use cases, telephony speech, and YouTube. The data also includes accented speech from multiple English locales \cite{sainath2020streaming}. The total amount of training data is $\sim$400k hours. We augment this training set by simulating noisy conditions at signal-to-noise ratio (SNR) between 0 and 30 dB, with the average SNR of 12 dB, and reverberation times ($T60$) between 0 and 900 msec, with an average $T60$ time of 500 msec \cite{kim2017mtr}. We also resample input utterances to have roughly equal number of $8k$~Hz and $16k$~Hz utterances. SpecAug \cite{park2019specaugment} and random state sampling \cite{narayanan2019longform} are used as regularizations.

As evaluation sets, we use short-form and long-form test sets. As short-form test sets, we use $\sim$12k anonymized and hand-transcribed utterances that represent Google voice-search (VS) and non-voice-search (NVS) traffic. NVS is relatively simpler than VS, since it has fewer rare-words, but contains utterances that are, on average, twice as long as those in VS. To create the long-form test set, passages containing at least 200 words from a novel are synthesized using a parametric text-to-speech system \cite{gonzalvo2016recent} (T-AB). T-AB contains $\sim$400 utterances, with an average duration of 65 seconds.

\subsection{Model architecture}

We use domain-id based multidomain models \cite{sainath2020streaming} in this work. As features, 128 dimensional log-Mel features are computed on a 32 msec window, with 22 msec overlap between neighboring windows. Features from 4 contiguous frames are stacked to form a 512 dimensional input feature, which is then sub-sampled by a factor of 3 along the time axis. A 16 dimensional one-hot domain-id vector is appended to this to form a 528 dimensional feature, which forms the input to the ASR model. 

We use multiple model architectures to show that cascaded encoders can be used irrespective of the underlying encoder type. The baseline model is an LSTM-based streaming RNN-T model. It consists of 8 encoder layers of unidirectional LSTMs. Each layer has 2048 cells, which are projected down to 640 output units. A time-reduction layer is used after the 2$^{nd}$ LSTM layer; the features output by the final layer of the streaming encoder is at 60 msec frame rate. The RNN-T decoder consists of 2 LSTM layers in the projection network. Similar to the encoder, each layer has 2048 units that is projected down to 640 output units. The output of the encoder and the projection network is passed to the joint network, which is a 1-layer neural net with 640 units. As output representation, we use a word-piece model with a vocabulary size of 4096 \cite{SchusterNakajima12}, and an embedding size of 128. In total, the model has $\sim$120M parameters 

When using conformers as the causal encoder in C-T, we use 17 layers. Each layer has 512-units, and uses 8 attention heads and a convolutional kernel of size 15. Note that all attentions are causal in C-T. Similar to RNN-T, the output features are stacked and subsampled by a factor of 2 after the 4$^{th}$ layer. The decoder still uses an LSTM-based prediction network and has the same structure as the baseline RNN-T model. The C-T model has $\sim$140M parameters.

All of our models are trained using Tensoflow \cite{abadi2016tensorflow} and the Lingvo toolkit \cite{ShenNguyenWuChenEtAl19}, on 8x8 slices of Tensor Processing Units \cite{jouppi2017datacenter}; we use a batch size of 4096 and Adam optimizer \cite{kingma2014adam}.
\section{Results}\label{sec:results}
\vspace{-0.05in}

We first present results when using LSTMs for building causal and non-causal layers. So in the subsequent subsection, we show how conformer layers can be used to limit the amount of future context the model has access to. Finally, we show how cascaded encoders can be used to improve performance when there are latency constraints for the first pass model. 

\subsection{Cascaded Encoders RNN-T model}\label{sec:rnnt-results}

\begin{table}[h]
  \begin{center}
  \begin{tabular}{lc|cccc}
    Models        && VS & NVS & T-AB \\ \hline
    RNN-T         && $5.9$ & $3.2$ & $4.5$ \\
    \multirow{2}{*}{CASE RNN-T}  & \multicolumn{1}{|c|}{C}  & $5.8$ & $3.2$ & $4.5$ \\
                                 & \multicolumn{1}{|c|}{NC} & $5.1$ & $2.6$ & $3.8$ \\
    BiDi RNN-T    && $4.8$ & $2.5$ & $3.8$ \\
    \hline
    LAS        && $5.2$ & $3.1$ & $20.5$ \\
    RNN-T + LAS Rescore  && $5.5$ & $3.0$ & $4.7$ \\
    \hline
    LAS-Delib  && $4.9$ & $3.1$ & $49.9$ \\
    RNN-T + LAS-Delib Rescore  && $5.3$ & $2.8$ & $4.7$ \\   
    \hline
  \end{tabular}
  \end{center}
  \vspace{-0.05in}
  \caption{WERs for baseline (RNN-T) and cascaded encoders RNN-T (CASE RNN-T) models. C and NC stand for causal and non-causal modes of CASE RNN-T. Also shown are results using bidirectional \mbox{RNN-T}, and LAS and LAS-deliberation two-pass models used to decode or rescore hypotheses.}
  \label{table:rnnt-results}
\end{table}

Tab.~\ref{table:rnnt-results} shows results using the RNN-T baseline, and the model that uses LSTM-based cascaded encoders (CASE RNN-T). The causal encoder in CASE RNN-T model has the same architecture as the baseline RNN-T. The non-causal encoder is a 2-layer bidirectional LSTM, with $\sim$10M parameters. Compared to the baseline RNN-T, CASE RNN-T in causal mode obtains very similar WERs. But when used in the non-causal mode, it improves WER on VS by 14\% and NVS by 19\%. On the long-form T-AB set, WER improves by 16\%. A fully bidirectional RNN-T model obtains similar WERs as the non-causal CASE RNN-T on NVS and T-AB sets, but improves VS WER by another 6\% relative. The results show that even with 2 layers of bidirectional LSTMs we can obtain WERs that are almost as good a fully bidirectional model using cascaded encoders.

The table also shows results using LAS-based \cite{sainath2019two} (LAS) and deliberation-based (LAS-Delib) \cite{hu2020deliberation} two-pass models models. Performance on VS improves compared to the RNN-T baseline, when using LAS or LAS-Delib in decode mode or rescore mode (first-pass hypotheses generated by RNN-T are rescored in this mode). But only LAS-Delib outperforms CASE RNN-T by 4\%, when the latter is used in non-causal mode. And on the slightly longer NVS set, and the long-form T-AB set, non-causal CASE RNN-T outperforms LAS and LAS-Delib. As expected, on T-AB, LAS and LAS-Delib marginally worsens WERs when used in rescore mode, and performs poorly when used in decode-mode. The results clearly show the advantages of CASE RNN-T over LAS and LAS-Delib.
\vspace{-0.06in}
\subsection{Limited right context}\label{sec:rnnt-conformer-results}
\vspace{-0.1in}
\begin{table}[h]
  \begin{center}
  \begin{tabular}{lc|c|ccc}
    \multirow{2}{*}{Models}    && Context & \multirow{2}{*}{VS} & \multirow{2}{*}{NVS} & \multirow{2}{*}{T-AB} \\ 
    {}        && (sec)   &  {}  & {}      & \\
    \hline
    RNN-T     && 0.0  & $5.9$ & $3.2$ & $4.5$ \\
    C-CASE RNN-T & \multicolumn{1}{|c|}{C} & 0.0  & $6.0$ & $3.2$ & $4.6$ \\ 
    \multicolumn{1}{r}{+ 2-layer conf.} & \multicolumn{1}{|c|}{NC} & 5.0  & $4.8$ & $2.5$ & $3.8$ \\
    \multicolumn{1}{r}{+ 3-layer conf.} & \multicolumn{1}{|c|}{NC} & 2.9  & $4.7$ & $2.6$ & $3.6$ \\
    \multicolumn{1}{r}{+ 4-layer conf.} & \multicolumn{1}{|c|}{NC} & 15.8 & $4.4$ & $2.4$ & $3.3$ \\
    \hline
    C-T         && 0.0  & $5.6$ & $3.1$ & $4.0$ \\
    \multirow{2}{*}{CASE C-T} & \multicolumn{1}{|c|}{C} & 0.0  & $5.8$ & $3.1$ & $4.1$ \\
                              & \multicolumn{1}{|c|}{NC}& 5.0  & $4.7$ & $2.7$ & $3.6$ \\
    \hline
  \end{tabular}
  \end{center}
  \vspace{-0.1in}
  \caption{WERs for RNN-T, C-T, cascaded encoders RNN-T when using conformer layers in the non-causal encoder (C-CASE RNN-T) or in both causal and non-causal encoders (CASE C-T). For C-CASE RNN-T, also included are WERs when the number of additional conformer layers and the amount of right context are varied. C stands for causal, and NC for non-causal modes.}
  \label{table:rnnt-conformer}
\end{table}

In order to limit the amount of future context the model sees, we use conformer layers in the non-causal encoder, while continuing to use LSTMs in the causal encoder (\mbox{C-CASE RNN-T}). Each conformer layer, in this setting, has 640 units to match the LSTM layers, and adds $\sim$10M additional parameters. The results are shown in Tab.~\ref{table:rnnt-conformer}. Switching the non-causal LSTM layers with conformer layers affect WERs of the causal cascaded encoders model only marginally. Running the same model in non-causal mode yields better WER improvements compared to using bidirectional LSTMs. Using 2 additional layers of conformers with 5.0 seconds of right context yields a relative WER improvement of 19\% on VS, 22\% on NVS, and 16\% improvement on T-AB. Using 3 layers with just 2.9 seconds of right context gives 20\% improvement on VS, 19\% on NVS, and 20\% improvement on T-AB. If we further increase the number of conformer layers to 4, and give the model 15.8 seconds of right context, VS, NVS and T-AB WERs improve by 25\%, 25\%  and 27\%, respectively. 

Also shown in Tab.~\ref{table:rnnt-conformer} are results when a conformer based encoder is also used for the streaming encoder. In line with the results in prior work \cite{gulati2020conformer}, C-T improves over RNN-T by 5\% on VS, 3\% on NVS and 11\% on T-AB. When using conformer-based cascaded encoders (\mbox{CASE C-T}) and decoding in the causal mode, performance on VS degrades by 4\% relative on VS, and similar WERs are obtained on the remaining sets. In the non-causal mode, CASE C-T obtains relative improvements of 16\% on VS, 13\% on NVS and 10\% on T-AB compared to C-T. Overall, conformer-based cascaded encoders outperform LSTM-based models in both causal and non-causal mode. 

These results, together with those in the previous subsection, show that we can easily combine multiple architectures for causal and non-causal layers in cascaded encoders model, thereby providing a flexible framework for building a unified streaming and non-streaming model.
\subsection{Faster first-pass}
Tab.~\ref{table:fastemit} shows results when the causal encoder in C-T is trained with FastEmit for improved partial latency, as described in Sec.~\ref{sec:fastemit}. Apart from WER, the table also shows the following latency metrics on the VS test set: 90$^{th}$ percentile latency from end-of-speech to final endpointing (EP90) and 90$^{th}$ percentile latency from end-of-speech to first correct partial hypothesis (PR90). PR90 measures how quickly the model predicts the correct hypothesis. As shown, when the cascaded encoders model (CASE C-T) is run in causal model, WER deteriorates on VS by 2\%, PR90 worsens by only 30 msec, and EP90 remains unchanged. When run in the non-causal model, which uses 2 additional conformer layers with 5 seconds of right context, WER improves by 21\% relative. The results show that the cascaded encoders can be used in a setting where streaming ASR is optimized for low latency, but the final result is generated via the non-streaming encoder that provides large performance gains over the causal model.
\begin{table}[h]
  \begin{center}
  \begin{tabular}{l|c|cccc}
    Models         & Context & VS & EP90 & PR90\\ 
    {}             & (sec)   & {}   & (msec) & (msec)\\
    \hline
    C-T             & 0   & $5.8$ & $660$ & $90$ \\
    CASE C-T        & 0   & $5.9$ & $660$ & $120$ \\
    + 2-layer conf. & 5.0 & $4.6$ & --    & -- \\
    \hline
  \end{tabular}
  \end{center}
  \vspace{-0.1in}
  \caption{WERs for baseline and cascaded RNN-T when using conformer-transducer and FastEmit.}
  \label{table:fastemit}
\end{table}
\vspace{-0.1in}
\section{Conclusion}
\label{sec:conclusion}
\vspace{-0.05in}
In this work, we presented cascaded encoders -- a flexible framework for combining streaming and non-streaming encoder architecture. Together with an RNN-T decoder, cascaded encoders can be used to build a single model that can operate in streaming and non-streaming mode. Compared to LAS-based two-pass models that have been proposed in the past, cascaded encoders provides gains for both short-form and long-form speech, and has a simplified one-stage training process. Our results also show that when the first pass model is optimized to minimize the E2E model's latency, cascaded encoders used as a second pass non-causal model can recover the resulting loss in WERs of the first-pass model. Future work will focus on combining cascaded encoders and deliberation models, since they, respectively, provide future context in partly complimentary acoustic and label spaces. Another interesting direction is to explore strategies for speeding up cascaded encoders when they are used as second pass model -- either by using rescoring instead of decoding, or by using decoders that use fewer computes than a typical LSTM-based RNN-T decoder.
\vspace{-0.1in}
\section{Acknowledgements} \label{sec:ack}
\vspace{-0.05in}
We thank Anmol Gulati, Bo Li, Ke Hu, Wei Han, and Yanzhang He for useful discussions and suggestions.


\bibliographystyle{IEEEbib}

\bibliography{ref}

\end{document}